\documentstyle[prl,aps,twocolumn]{revtex}
\begin{document}
\voffset 0.5in
\draft
\wideabs{
\title{Density-functionals not based on the electron gas:
Local-density approximation for a Luttinger liquid}
\author{N.~A. Lima, M.~F. Silva, and L.~N. Oliveira}
\address{Departamento de F\'{\i}sica e Inform\'atica,
Instituto de F\'{\i}sica de S\~ao Carlos,
Universidade de S\~ao Paulo,\\
Caixa Postal 369, 13560-970 S\~ao Carlos, SP, Brazil}
\author{K. Capelle}
\address{Departamento de Qu\'{\i}mica e F\'{\i}sica Molecular,
Instituto de Qu\'{\i}mica de S\~ao Carlos,
Universidade de S\~ao Paulo,\\
Caixa Postal 780, S\~ao Carlos, 13560-970 SP,
Brazil}
\date{\today}
\maketitle
\begin{abstract}
By shifting the reference system for the local-density approximation (LDA) 
from the electron gas to other model systems one
obtains a new class of density functionals, which by design account for
the correlations present in the chosen reference system. 
This strategy is illustrated by constructing an explicit LDA for the 
one-dimensional Hubbard model. While the traditional {\it ab initio} LDA is 
based on a Fermi liquid (the electron gas), this one is based on a Luttinger 
liquid. First applications to inhomogeneous Hubbard models, 
including one containing a localized impurity, are reported.
\end{abstract}

\pacs{71.15.Mb, 71.10.Pm, 71.10.Fd, 71.27.+a}
}

\newcommand{\be}{\begin{equation}}
\newcommand{\ee}{\end{equation}}
\newcommand{\bea}{\begin{eqnarray}}
\newcommand{\eea}{\end{eqnarray}}
\newcommand{\bi}{\bibitem}

\newcommand{\ep}{\epsilon}
\newcommand{\s}{\sigma}
\newcommand{\p}{{\bf \pi}}
\newcommand{\r}{({\bf r})}
\newcommand{\rp}{({\bf r'})}
\newcommand{\rpp}{({\bf r''})}

\newcommand{\ua}{\uparrow}
\newcommand{\da}{\downarrow}
\newcommand{\la}{\langle}
\newcommand{\ra}{\rangle}
\newcommand{\dg}{\dagger}

Density-functional theory (DFT) \cite{hkks} is the basis of almost all of 
todays electronic-structure theory, and much of materials science and quantum 
chemistry. 
Many-body effects enter DFT via the exchange-correlation ({\it xc}) functional,
which is commonly approximated by the local-density approximation (LDA) 
\cite{hkks}. The essence of the LDA is to locally approximate the {\it xc} 
energy of the inhomogeneous system under study by that of the homogeneous 
electron gas. This electron gas plays the role of a {\it reference system},
whose correlations are transfered by the LDA into the DFT description of 
the inhomogeneous system. The most popular improvement upon the LDA are
generalized gradient approximations \cite{gga}, whose basic philosophy
is to abandon the requirement of homogeneity of the reference system. This
system, however, is normally still the interacting electron gas \cite{gga}.

In the present paper we propose to explore a different paradigm for the 
construction of novel density functionals: instead of sticking to the 
electron gas as a reference system, and abandoning homogeneity, it may
sometimes be advantageous to do the reverse: stick to homogeneity
(and thus to the LDA) but abandon the electron gas as a reference system. 
The new reference system is chosen such that it accounts for the
correlations present in the inhomogeneous system under study.

The only requirement for the reference system is that in the absence of 
any spatially varying external potential its {\it xc} energy must be known 
exactly or to a high degree of numerical precision.
Besides the electron gas (or Jellium model) there are many other
physically interesting model systems that satisfy this criterium.
Most notably among these is a large class of low-dimensional models which can 
be solved {\it exactly} by Bethe Ansatz (BA) techniques or bosonisation 
(in one dimension, e.g., the repulsive and the attractive Hubbard model, 
the hard-core Fermi and Bose gases, the Heisenberg, the supersymmetric 
t-J, and the Tomonaga-Luttinger model \cite{schlottmann,voit}). 
The solutions to these models in the homogeneous case can  
be used instead of the electron gas to construct LDA functionals
that can then be applied to study these models also in inhomogeneous 
situations. The main advantage offered by a DFT treatment of such models is 
the gain in simplicity that arises from mapping the inhomogeneous 
interacting many-body system onto a noninteracting auxiliary system, which 
is diagonalized much more easily.

Below we implement these ideas for the one-dimensional Hubbard model (1DHM),
by constructing an LDA based on the
exact Bethe Ansatz solution of Lieb and Wu \cite{bethe}.
A DFT treatment of the Hubbard model has been pioneered by Gunnarsson and
Sch\"onhammer in Ref.~\cite{gs}, but the LDA-type functional they proposed
has in practice often led to disapointing results \cite{runge}, and
was criticized as not being a proper LDA since it was not
based on the exact solution of a homogeneous reference system \cite{sham}.
As a consequence, more complicated approximation schemes, 
such as self-interaction corrections, are often employed
\cite{sic}. An attempt to base a proper LDA for the Hubbard model
on the BA was made in Ref.~\cite{sgn}, but the formulation
of that work has not been widely applied, probably because no 
explicit expression for the resulting {\it xc} functional was provided. 

Motivated by our above analysis of the possibility of switching
reference systems for the LDA we construct, in the present paper, an
explicit and simple BA {\it xc} functional and apply it to a variety of 
inhomogeneous Hubbard models, among them an impurity model for previously 
unattainable system sizes.
This functional, denoted the BA-LDA, has built into it the
Luttinger-liquid correlations present in the 1DHM \cite{schlottmann,voit},
in the same way in which the conventional LDA has built into it the
Fermi-liquid correlations present in the electron gas.

The Hamiltonian for the homogeneous 1DHM is, in standard notation,
\be
\hat{H}=-t\sum_{\langle ij \rangle,\sigma} c_{i\sigma}^\dagger c_{j\sigma}
+U\sum_i c_{i\ua}^\dagger c_{i\ua}c_{i\da}^\dagger c_{i\da}.
\label{1dhm}
\ee
Here $t$ is the kinetic energy (in the following taken to be the unit of 
energy) and $U$ the interaction (considered a fixed parameter, characterizing 
the Hamiltonian).
To construct an LDA we first develop a parametrization for the total energy 
per site, as a function of $U$ and $n$ (the filling factor, a constant in the 
homogeneous case). Our parametrization interpolates analytically between
three limiting cases in which explicit results can be
extracted from the Bethe-Ansatz solution \cite{schlottmann,bethe}: 
(i) $U\to 0$ and any $n \leq 1$, 
(ii) $U\to \infty$ and any $n \leq 1$, and 
(iii) $n=1$ and any $U$ \cite{phtrafofootnote}.
The expression
\be
e^{BA}(n,U) = 
-\frac{2\beta(U)}{\pi} \sin\left(\frac{\pi n}{\beta(U)}\right),
\label{ehom}
\ee
where $\beta(U)$ is an ($n$ independent) number 
which is determined for any given value of $U$ from
\be
-\frac{2\beta}{\pi} \sin\left(\frac{\pi}{\beta}\right) =
-4\int _0^\infty dx
\frac{J_0(x)J_1(x)}{x[1+\exp(Ux/2)]},
\label{betadet}
\ee
and $J_0$ and $J_1$ are zero and first order Bessel functions,
recovers all three limits:
The right-hand side of Eq.~(\ref{betadet}) is the exact BA expression for the 
total energy at half filling. The parameter $\beta$ is thus determined such
that at $n=1$ Eq.~(\ref{ehom}) becomes exact. 
On the other hand, Eq.~(\ref{ehom}) is already of the algebraic form of
the exact results for the limits $U=0$ and $U\to \infty$, in which
$\beta=2$ and $\beta=1$, respectively. 
In these limits the integral in Eq.~(\ref{betadet}) can be calculated 
analytically, and one indeed recovers these values for $\beta(U)$
\cite{betafootnote}.

Eq.~(\ref{ehom}) with $\beta(U)$ determined from (\ref{betadet}) is thus exact 
in the three situations mentioned above. In order to check whether these 
equations provide a reasonable approximation also between these limits, we 
have numerically solved the Lieb-Wu integral equations for the full Bethe 
Ansatz solution \cite{bethe} and compared the resulting total and $xc$ energies
with the one obtained from Eqs.~(\ref{ehom}) and (\ref{betadet}). We find that 
both agree to within at most a few percent, even for values of the parameters
far away from the exact limiting cases. This is illustrated in the inset
of Fig.~\ref{fig1}. 

In order to extract the exchange-correlation energy $e^{BA}_{xc}(n,U)$ from 
the total-energy expression (\ref{ehom}), one follows the usual prescription 
of DFT\cite{hkks} and subtracts the Hartree energy and the noninteracting 
kinetic energy. This latter energy is simply given by substituting $\beta=2$ 
in Eq.~(\ref{ehom}). In the spirit of the usual electron gas LDA we can then
construct an LDA for the 1DHM, i.e.,
\be
E_{xc}^{BA-LDA}[n_i;U]=\sum_i e_{xc}^{BA}(n,U)|_{n\to n_i},
\label{lda}
\ee
where $n_i=\sum_\sigma\langle c_{i\sigma}^\dagger c_{i\sigma} \rangle$.
Given this expression for the {\it xc} functional, ground-state properties of 
Hubbard models subject to a wide spectrum of inhomogeneities can be calculated 
from DFT.
Note that in such a calculation Eq.~(\ref{betadet}) must be solved only once
for any given value of $U$, i.e., determination of $\beta$ takes place outside 
the self-consistency cycle of DFT.

As a first numerical example we apply the BA-LDA to a finite 1DHM with 
open boundary conditions, and calculate its ground-state energy as a function 
of the number of sites $L$. In Fig.~\ref{fig1} we compare our results with
those obtained from exact (Lanczos) diagonalization of the same system. We
see that around $L=7$ there is a crossover between exact and approximate
data points (separately for even and odd values of $L$). After that, the
deviation between both sets of data saturates to a value near the intrinsic
error of the interpolation (\ref{ehom}), indicated by the error bar.
As an example for a truly inhomogeneous system we now add an on-site potential 
$\sum_{i\s} v_i c_{i\s}^\dagger c_{i\s}$
to the Hamiltonian. Our results for a binary potential (with $v_i=-1$ on the 
odd sites and $v_i=+1$ on the even ones) are displayed and compared with exact 
diagonalization in Fig.~\ref{fig2}.

We have performed similar calculations also for several other values of 
$U$ and other external potentials $v_i$. Our conclusions from these
calculations are: (i) The accuracy of the BA-LDA total energy is typically of 
the order of a few percent, and much better than that near crossovers and
near the limits at which equation (\ref{ehom}) exactly represents the
underlying homogeneous reference data ($U=0$, $U\to \infty$, $n=1$).
(ii) Unlike traditional methods, the quality of the BA-LDA does not 
deteriorate as $L$ increases, and the computational effort associated 
with it is that of diagonalizing a noninteracting system. Fully self-consistent
calculations for systems with tens of thousands of sites are thus possible.
This is a unique, and rather desirable, feature of the BA-DFT, as compared 
to traditional methods.
 
Interestingly, when one of the site occupation numbers comes very close 
to $1$ (typically within less than $5\times 10^{-3}$), the self-consistency 
cycle associated with the Hubbard model Kohn-Sham equations does not 
necessarily converge.
In the homogeneous case $n=1$ (half filling) marks the Mott metal-insulator
transition associated with the opening of a gap in the energy spectrum
\cite{schlottmann,voit,bethe}. Whenever in a metallic system one of the
site occupation numbers comes close to $1$, a local approximation, such as
the BA-LDA, thus treats the system at that site as if it were an insulator, 
in spite of the fact that the metallic (Luttinger liquid) correlations are 
very different from those of the Mott insulator. 
The 1DHM thus constitutes a theoretical laboratory in which the band-gap 
problem of DFT can be studied \cite{gs,sgn}. Results obtained with the 
present functional will be reported in a forthcoming publication.

After these preliminary investigations we now consider a case that
illustrates the full power of the BA-LDA approach: a localized impurity. 
For this kind of system traditional approaches face the problem of
slow convergence to the 
thermodynamic limit. Different types of impurities in the Hubbard model 
have been studied in the literature by various techniques \cite{gs,inhom}. 
Here we model the impurity by choosing $v_I=-1$ at the impurity site and 
$v=0$ everywhere else, so that electrons will be dragged to the impurity 
site. The density distribution for the 201-site system is displayed in 
Fig.~\ref{fig3}, while the convergence to the thermodynamic limit is
illustrated in Fig.~\ref{fig4}, which also contains results obtained
for a much more attractive impurity with $v_I=-10$. 

The difference between open and periodic boundary conditions becomes small 
only when the system size $L$ exceeds the damping length of the Friedel 
oscillations originating at the surface in the open case.
To bring out clearly the effect of the impurity, regardless of the choice 
of boundary conditions, we have, in both figures, subtracted the results 
for the same boundary conditions without the impurity. What remains are 
the Friedel oscillations originating at the impurity. These oscillations 
on their own significantly slow down the convergence of the total energy 
to the thermodynamic limit: Fig.~\ref{fig4} shows that, as expected in the
thermodynamic limit, the impurity-added contribution to the total energy
scales linearly with the impurity concentration $1/L$. 
The impurity problem thus illustrates an area in which the BA-LDA can be 
useful, since systems of the size required to approximate the thermodynamic 
limit to within a percent or better are hard to study with traditional methods,
in particular for periodic boundary conditions. 

For small systems, where exact diagonalization is possible, we also
compared the density distributions obtained from the BA-LDA with the exact 
ones. Both agree quantitatively. For larger systems one can use the BA-LDA
results to study the asymptotic algebraic decay of the oscillations. For the 
case depicted in Fig.~\ref{fig3} we find, for example, that the oscillations
decay as $1/x^\gamma$, where $x$ is the distance to the impurity site and 
$\gamma(U=6)=1.30$. This exponent is a nonuniversal 
(interaction-dependent) parameter characteristic for the impurity system. 
By repeating this calculation for other values of $U$ we obtain, e.g., 
$\gamma(U=4)=1.25$, $\gamma(U=2)=1.20$, and $\gamma(U=0)=1.0$.

For the BA-LDA the limit on the size of the systems which can be treated is 
much less restraining than for traditional methods, since one must diagonalize 
only a noninteracting (Kohn-Sham) Hamiltonian.
For {\it small} systems the accuracy attained is clearly inferior 
to density-matrix renormalization group 
\cite{dmrg} or Quantum Monte Carlo \cite{qmc} methods.
This situation parallels the one in which DFT finds itself in 
{\it ab initio} calculations: Practical applications of {\it ab initio} DFT 
usually do not lead to high accuracy.
Band structures, for example, are more accurately calculated using
GW\cite{gw}, and properties of small molecules are better obtained from
CI\cite{ci}. However, these are computationally expensive methods
that place great demands on ones resources and are not easily applicable
to complex and/or large systems (e.g. molecules with more than a few
atoms). 
The power of {\it ab initio} DFT arises from the relative facility with which 
it is applied to large and complex systems. 
In the same spirit, DFT using generalized LDA's may provide a useful 
alternative to traditional methods for correlated models with a
large number of sites. In order to further explore these possibilities,
we are currently applying the BA-LDA to the Mott insulating phase of the
1DHM and to a more detailed study of the Friedel oscillations around an
impurity. Results will be reported in forthcoming publications. 

{\bf Acknowledgments}
This work was sup\-por\-ted by FAPESP and CAPES. KC thanks 
A.~J.~R.~da Silva for valuable discussions.

\begin{figure}
\caption{Main figure: Total energy calculated from BA-LDA (open squares) and 
exact diagonalization (full circles) versus system size, for a 1DHM with 
open boundary conditions and $U=6$. For even $L$ we take $N=L/2$ (so that
$n=N/L=1/2$, corresponding to quarter filling). For odd $L$ we take
$N=(L-1)/2$. The error bar at $L=14$ represents the intrinsic error of the
parametrization (\ref{ehom}), as estimated from the inset. For $L=14$ the 
BA-LDA data agree with the exact ones within this error.
Inset: Exchange-correlation energy of the infinite 1DHM as a function of $n$ 
for several values of $U$. The full curves represent our 
parametrization (\ref{ehom}) with (\ref{betadet}), and the symbols represent
values obtained from numerical solution of the Lieb-Wu BA integral equation
for $U=3,6$, and $9$. For $U=0$ and $U=\infty$ the parametrization is exact.
}
\label{fig1}
\end{figure}

\begin{figure}
\caption{Main figure: Total energy of the 1DHM with a binary site potential 
and periodic
boundary conditions, as described in the main text, for $U=6$ and $n=1/2$.
Open squares: BA-LDA, full circles: exact diagonalization. Inset: exact
$xc$ energy and LDA $xc$ energy, evaluated at exact and LDA densities,
respectively. The absolute errors in the LDA $xc$ energy and total energy 
are different because in the LDA the Hartree, external potential and 
noninteracting kinetic components of the total energy are also evaluated 
at the LDA density, not at the exact one. However, the $xc$ energy is the 
dominating source of error in the total energy and the relative error in 
the total energy is considerably smaller than that in its $xc$ component.}
\label{fig2}
\end{figure}

\begin{figure}
\caption{Density distribution for a system with $L=201$ sites,
$N=101$ electrons, $U=6$, and a localized impurity with $v_{imp}=-1$. 
Main curve: periodic boundary conditions. Inset: open boundary conditions.
The density of the same system without impurity has been subtracted to
display clearly the impurity-added contribution.}
\label{fig3}
\end{figure}

\begin{figure}
\caption{Total energy calculated from BA-LDA for the impurity model of
Fig.~\ref{fig3}, with open (full symbols) and periodic (empty symbols)
boundary conditions, plotted as a function of the inverse system size.
Impurity strength $v_I=-1$ (upper curve), and $v_I=-10$ (lower curve).}
\label{fig4}
\end{figure}

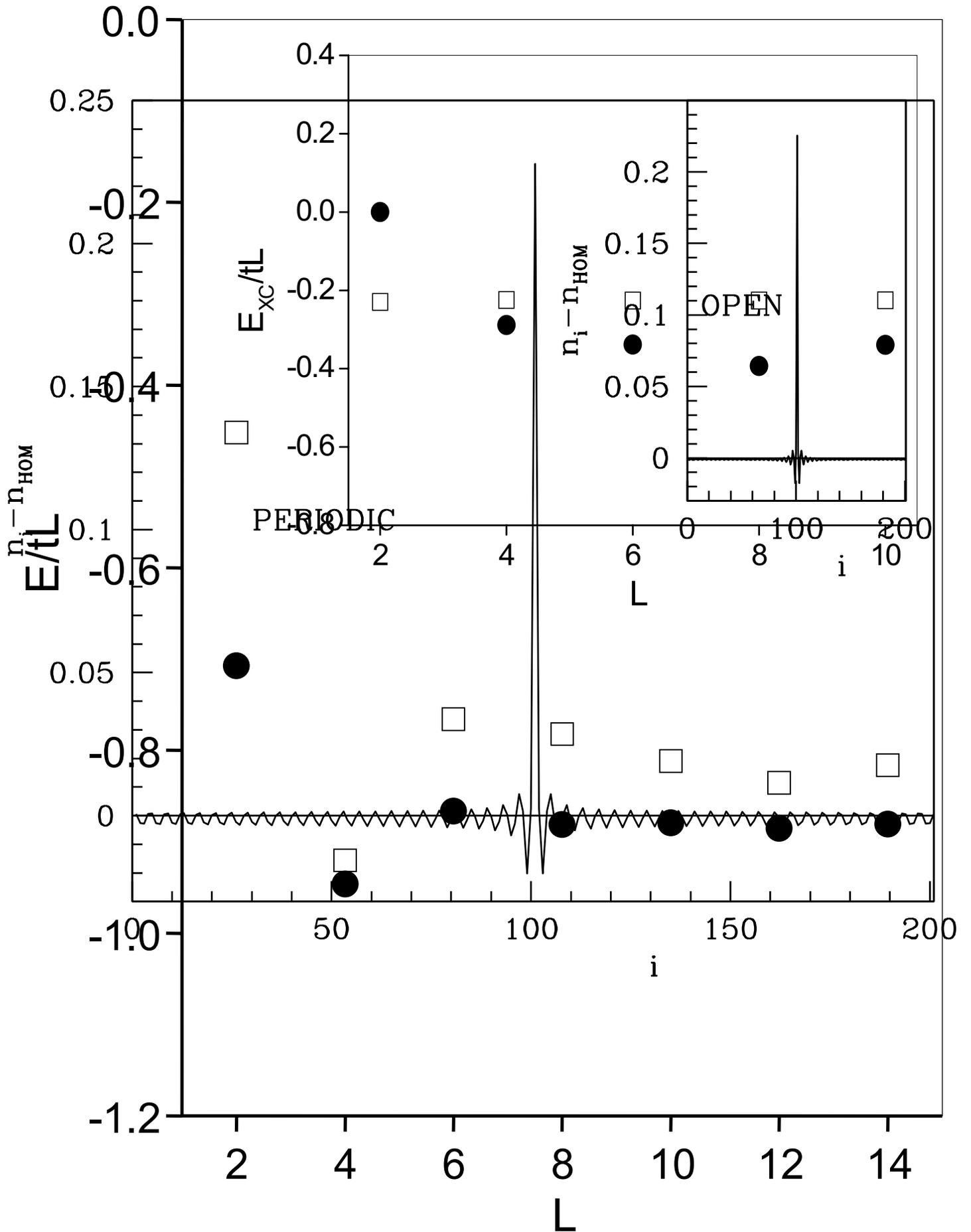
\end{document}